\documentstyle[aps,prl,multicol,epsf]{revtex}
\def\be{\begin{equation}}
\def\ee{\end{equation}}

\def\bea{\begin{eqnarray}}
\def\eea{\end{eqnarray}}

\begin{document}

\title{Presence of energy flux in quantum spin chains: \\
An experimental signature}
\vspace {1truecm}

\author{Z. R\'acz}
\address{{}Institute for Theoretical Physics,
E\"otv\"os University,
1117 Budapest, P\'azm\'any s\'et\'any 1/a, Hungary}
\date{\today}
\maketitle
\begin{abstract}
Using the $XXZ$ model for the description of  one-dimensional
magnetic materials  we show that  
an energy flux, $j_E$, produces a shift, 
$\delta k\sim \sqrt{j_E}$, in the characteristic wavenumber of 
the spin-spin correlations. 
We estimate $\delta k$ for a realistic experimental setup 
and find that it is measurable in inelastic neutron scattering experiments.

\vspace{0.1truecm}
\end{abstract}

\date{\today}

\maketitle
\begin{multicols}{2}
\narrowtext
\section{Introduction}

The problem of thermal transport in one-dimensional systems has  been
much investigated, the main goal being to derive Fourier's heat
law. Analytical and numerical studies of a number of classical
lattice-dynamical models indicate that the condition for Fourier's
law to hold is the presence of strong nonlinearities
i.e. nonintegrability (chaoticity) of the dynamics
\cite{Fourier}. Although quantum systems have been less studied, it
appears that similar considerations apply to quantum spin chains as
well\cite{Saito}. 

Integrable systems, on the other hand, show anomalous 
thermal transport. No internal thermal gradient is formed 
in a harmonic crystal \cite{Lebo} or in a transverse Ising chain \cite{Saito}
and, as a consequence, 
the energy (heat) flux is not proportional to the temperature gradient 
inside the sample. The origin of this anomaly may be the fact that the energy 
current in integrable systems often emerges as an integral of motion 
which automatically yields anomalous thermal transport 
coefficients \cite{Zotos}. 

The flat temperature profile in the presence of energy current 
is an intriguing feature of integrable systems. In effect, 
it points to the  existence  of a homogeneous state 
carrying  finite energy current. In this paper, 
we shall explore the experimentally 
measurable properties of such a state by studying the $XXZ$ spin chain
in the presence of an energy current. The most spectacular feature of
such a state is the incommensurability of magnetic excitations. Namely,  
in the presense of energy flow $j_E$,  the characteristic wave vector 
is shifted from its antiferromagnetic value $\pi$ by the 
amount $\delta k\sim \sqrt{j_E}$. 
Since there are quite a few  well established realizations of 
quasi-one-dimensional Heisenberg chains \cite{1DHeis1} (e.g. 
${\rm KCuF_3}$ \cite{kcuf3}, ${\rm Cs_2CoCl_4}$ \cite{cs2cocl4},
Copper Benzoate \cite{cubenz}, ${\rm Sr_2CuO_3}$, ${\rm Cs_2CuCl_4}$
\cite{cscucl}), we believe that the predicted changes in the dynamical
correlation functions bear direct experimental relevance.

The basic problem of constructing a state which carries an energy  
current is the nonequilibrium nature of that state. Even if we assume 
that the flat temperature profile means the existence of 
equilibrium, we still face a  problem that the value of
the established temperature is not known \cite{DRS}. We shall 
avoid this problem by restricting our calculation to zero temperature 
$(T=0)$ and assuming that the ground state correlations are robust enough
to survive at low temperatures. 

The construction of a
homogeneous state with energy current at $T=0$ can be done by
adding the energy current with a Lagrange multiplier 
to the $XXZ$ Hamiltonian and then finding the ground state.
Similar calculations have been carried out 
already for the transverse Ising and $XX$ chains 
\cite{{Antal1},{Antal2}} and, in a different context, for the 
$XXZ$ model \cite{{TsvelikXXZ},{FrahmXXZ}}. 
The new result we report is the calculation of an experimentally 
accessible parameter, namely the shift, $\delta k$, 
of the characteristic wavenumber in the spin-spin correlations as a
function of the energy current, $j_E$. 

Once we have $\delta k(j_E)$, we turn to a realistic experimental setup
and estimate $j_E$ flowing through a single spin chain which gives an 
estimate of $\delta k$. Our result shows that $\delta k$ is 
in the accessible range of an inelastic neutron scattering 
experiment.

\section{The model and the characteristic wave number}

The model we study is the spin-1/2 XXZ
chain defined by the Hamiltonian 
\begin{equation}
\hat H_{\rm xxz}={\cal J}\sum_\ell \left[\sigma^x_\ell\sigma^x_{\ell+1} +
\sigma^y_\ell\sigma^y_{\ell+1}+
\Delta \sigma^z_\ell\sigma^z_{\ell+1}\right]\quad ,
\label{H_XXZ}
\end{equation}
where the spins
$ \sigma^\alpha_\ell$ ($\alpha =x,y,z$)  are Pauli spin matrices at sites
$\ell=1,2,...,N$ of a one-dimensional periodic chain
($\sigma^\alpha_{N+1}= s^\alpha_1$). We shall use the 
parametrization $\Delta=\cos{\gamma}$ and consider 
only the `antiferromagnetic' region $0<\gamma<\pi/2$. In order
to impose a fixed energy current $j_E$ in the ground state, we add the
current operator to the Hamiltonian with a Lagrange multiplier
\be
\hat H = \hat H_{\rm xxz} + \lambda \hat{j}_E \label{mod}\ .
\ee
where
\bea
&\hat{j}_E=\frac{{\cal J}^2}{\hbar} \sum_\ell \sigma_\ell^z \Big[
\sigma_{\ell-1}^y\sigma_{\ell+1}^x-\sigma_{\ell-1}^x\sigma_{\ell+1}^y+
\nonumber\\
&\Delta (
\sigma_{\ell-2}^x\sigma_{\ell-1}^y-\sigma_{\ell-2}^y\sigma_{\ell-1}^x+
\sigma_{\ell+1}^x\sigma_{\ell+2}^y-\sigma_{\ell+1}^y\sigma_{\ell+2}^x)
\Big]\ .\label{currentop}
\eea
Importantly, $\hat{j}_E$ is an integral of motion, $[\hat{j}_E,\hat{H}]=0$,
thus indicating that i) the transport of energy is singular 
in this system \cite{Zotos} and ii) the states carrying fixed 
energy current can be obtained as stationary states of ${\hat H}_{\rm xxz}$. 

The $XXZ$ model can be described in terms of interacting fermions and 
has been solved using the Bethe Ansatz method.  
The same approach works 
in the presence of the driving term, $\lambda {\hat j}_E$, as well, and the 
solution has been given in \cite{TsvelikXXZ,FrahmXXZ}.
An interesting feature of the solution is 
that the system displays rigidity against the drive, namely the 
ground state supports a nonzero energy current, 
$\langle {\hat j}_E\rangle \equiv j_E\not=0$, only 
if the coupling $\lambda$ exceeds some critical value $\lambda_c(\gamma)$. 
As we are interested in fixed energy currents, we simply choose 
sufficiently large values of
$\lambda$. Furthermore, since in realistic situations  
$j_E$ turns out to be small, we concentrate on the region $\lambda\approx
\lambda_c(\gamma)$, in which case $j_E\propto (\lambda-\lambda_c)$. 

Once the energy current flows, an important restructuring takes place 
in the ground state. The single Fermi sea characterizing the 
ground state without current splits into {\sl two} Fermi seas as
shown in Fig.1 for the simple case of the $XX$ limit ($\Delta=0$)
where a free-fermion description applies. There are now four Fermi 
wave vectors, $\pm \pi/2$ and $\pi/2\pm \delta k$, and the 
structure of the ground state immediately implies that
there will be gapless excitations at wave vectors $0,\delta
k, 2\delta k, \pi-\delta k, \pi$ and $\pi+\delta k$, which is readily
confirmed by the exact solution at arbitrary $\Delta$. 

\begin{figure}[htb]
\vspace{3truecm}
\centerline{
        \epsfxsize=8cm
        \epsfbox{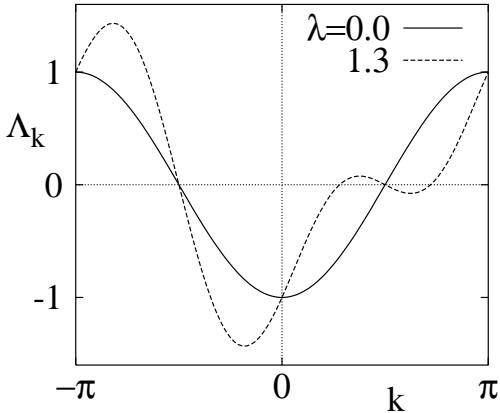}
           }
\vspace{-2.8truecm}
\caption{Single-particle fermionic spectrum in the $XX$ limit ($\Delta=0$)
of the ${\hat H}_{\rm xxz}+\lambda {\hat j}_E$ hamiltonian 
with and without energy current in the ground state
(dashed line, $\lambda =1.3$ and solid line, $\lambda =0$, respectively).
The energy is measured in units of $\cal J$ while the wave number, $k$,
is given in units of the inverse lattice spacing. The 
Fermi energy is zero independently of $\lambda$.}
\label{fig1}
\end{figure}
Thus an incommensurability characterized by 
$\delta k$ appears in the system. This can be seen readily in the 
ground-state correlations. Indeed, it has been shown  \cite{Antal2,Rakos} 
that, in the $XX$ limit, the longitudinal 
correlations for small $j_E\not= 0$ can be expressed in a scaling form 
\begin{equation}
\frac{\langle \sigma^x_\ell \sigma^x_{\ell+n}\rangle_{j_E\not= 0}}
{\langle \sigma^x_\ell \sigma^x_{\ell+n}\rangle_{j_E=0}} =
\Phi(\delta k\, n)
\label{xxcorrXY}
\end{equation}
where $\Phi(x\to 0)=1$ and the large argument asymptotics of 
the scaling function is given by 
\begin{equation}
\lim_{x\to \infty}\Phi(x)\sim \frac{1}{\sqrt{x}}(1+\cos{x}) \quad .
\label{Phi}
\end{equation}  
Since $\langle \sigma^x_\ell \sigma^x_{\ell+n}\rangle_{j_E=0}
\sim (-1)^n/\sqrt{n}$ equations (\ref{xxcorrXY},\ref{Phi}) imply that, as the 
current is switched on, the static structure factor 
develops additional peaks at $k=\pi\pm \delta k$ (as it will turn out, 
$\delta k$ is small thus it is better to speak about the $k=\pi$ peak 
developing shoulders for $j_E\not=0$).

In order to connect $\delta k$ to the current one determines both
$j_E$ and $\delta k$ through $\lambda$ and then eliminates 
the Lagrange multiplier. The expressions are simple for the $XX$ limit
\cite{Antal2}
\begin{equation}
j_E = \frac{{\cal J}^2}{2\pi \hbar}\left(1-\frac{1}{\lambda^2}\right)\quad ,
\quad \cos{\delta k}=\lambda^{-1} \quad \label{JXY}
\end{equation}
and, for small currents ($\lambda\ge\lambda_c=1$), they yield
\begin{equation}
\delta k =\sqrt{\frac{j_E}{j_E^{(1)}}}\quad .
\label{deltak}
\end{equation}
where a `natural unit' of the current, $j_E^{(1)}={\cal J}^2/h$, 
has been introduced. 

The above calculation can be carried out for any $0\leq  \Delta < 1$ 
and the result for $\delta k$ differs only in a prefactor of order unity 
\cite{esstsve}
\be
\delta k=\frac{2\gamma}{\pi\sin\gamma}\sqrt{\frac{j_E}{j_E^{(1)}}}
\quad .
\label{kj}
\ee
As one can see, the largest $\delta k$ is obtained in the 
$XX$ limit ($\gamma\to \pi/2$).

In principle, if $\delta k$ is large enough then the extra peaks 
at $\pi\pm \delta k$ should be observable as Bragg peaks in an elastic 
neutron scattering experiment. In practice, however, 
the incommensurate modulations of distinct 
spin chains are not correlated and, as a consequence, the 
delta function of the Bragg peak 
would spread out into a plain and the effect would be
unobservable.

\section{Structure factor and an experimental setup}

It is more promising to look for an experimental signature in an 
inelastic neutron scattering experiment where the excitations of the system 
are measured and no coherence among the chains is needed.
Taking into account the facts that, for $j_E=0$, 
most of the spectral weight
is concentrated on the region around the
antiferromagnetic wave vector $\pi$, and furthermore that, for 
$j_E\not= 0$, there are gapless 
excitations at wave vectors $\pi\pm\delta k$, 
we expect that the presence of the current manisfests itself
via the emergence of {\em
additional} inelastic peaks at wave vector $\pi\pm\delta k$. 
This expectation can be put on a more solid base by calculating 
the dynamic structure factor and examining the relative weights at
wave vectors $\pi$ and $\pi\pm\delta k$. 

The simplest case is 
again the $XX$ limit where the calculation of the time-dependent 
transverse correlation function, 
$\langle \sigma_n^z(t)\sigma_0^z(0)\rangle$, is straightforward.
There is, however, a principal problem at the outset of the calculation.
Namely, it is not clear whether the time-evolution of $\sigma_n^z(t)$
is governed by $\hat H_{\rm xx}$ or by $\hat H_{\rm xx}-\lambda \hat j^E$.  
We shall take the view that in reality the current-carrying state is
formed as a result of boundary conditions. Thus the local perturbation
caused by an incoming neutron evolves by the local hamiltonian i.e. 
by $\hat H_{\rm xx}$ \cite{notehxx}. Once 
$\langle \sigma_n^z(t)\sigma_0^z(0)\rangle$ is known
its Fourier transform in time and space gives the structure 
factor $S_{zz}(k,\omega)$ as shown in Fig.2.
 
\begin{figure}[htb]
\centerline{
        \epsfxsize=7cm
        \epsfbox{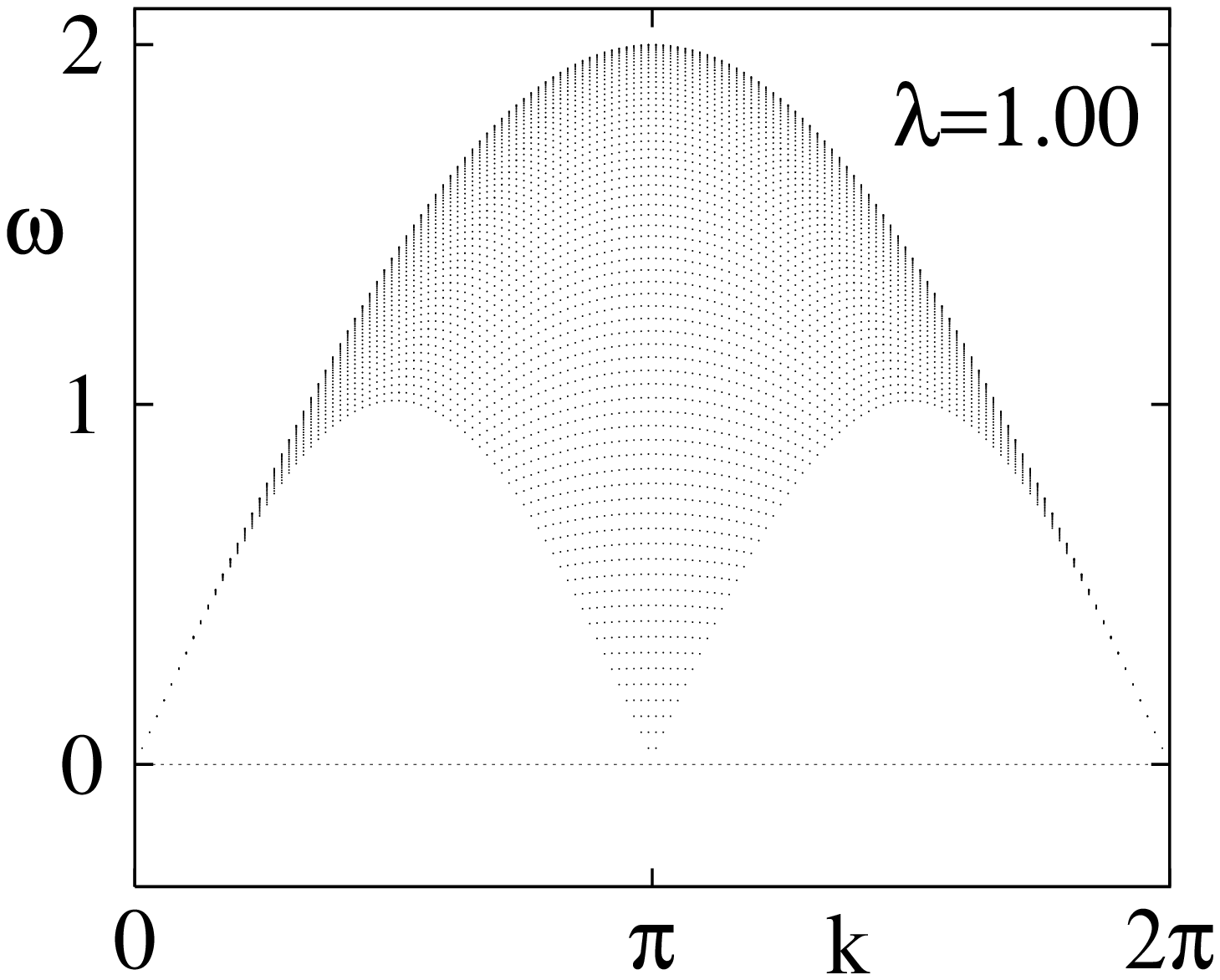}
           }
\vspace{0.3cm}
\centerline{
        \epsfxsize=7cm
        \epsfbox{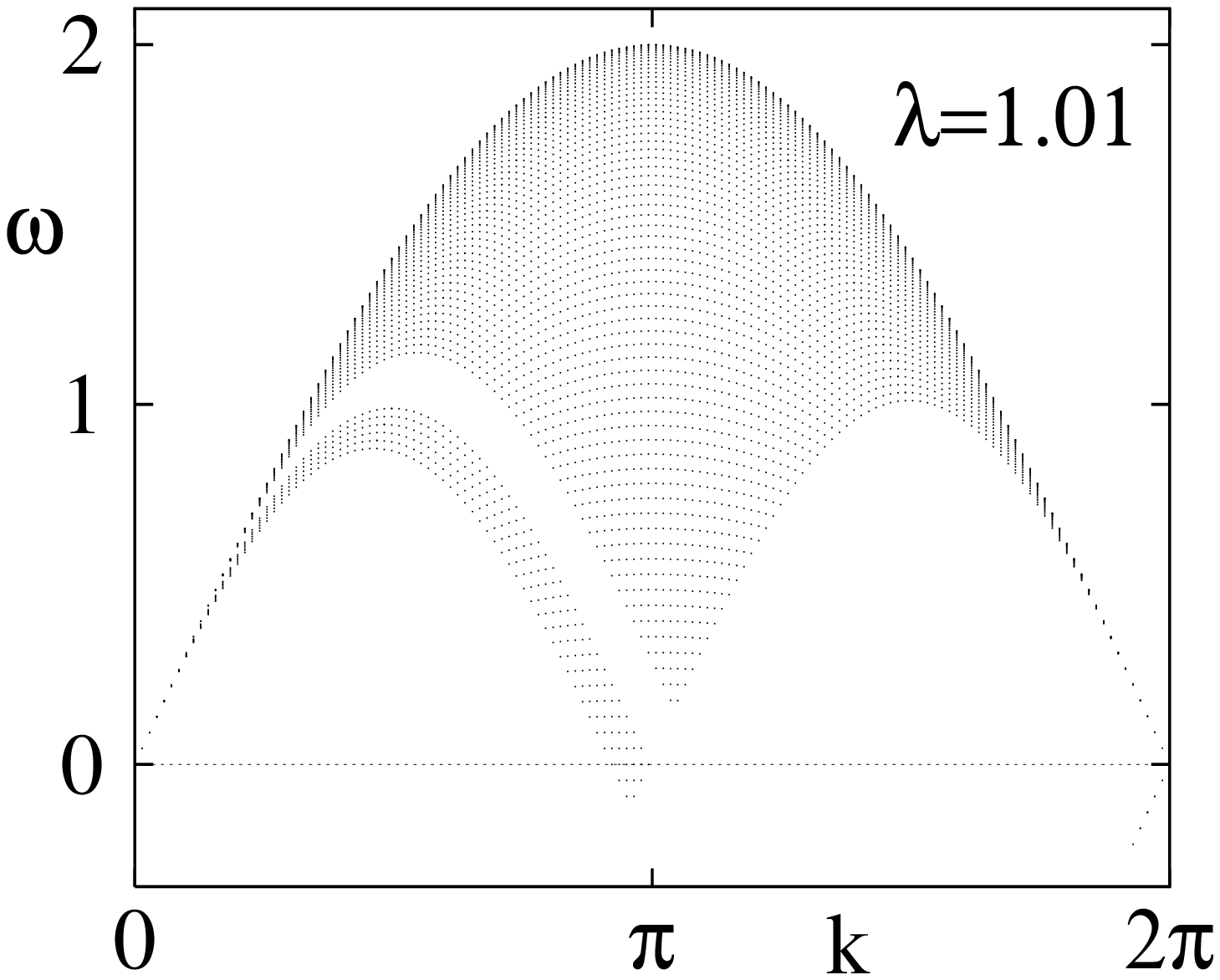}
           }
\vspace{0.2truecm}
\caption{The structure factors $S_{zz}(k,\omega)$ displayed 
on the wavenumber-frequency ($k-\omega$) plane for cases of (a)
$\lambda=1.00$ (no flux) and (b) $\lambda =1.01$. The unit of $\omega$ is
${\cal J}/\hbar$ while $k$ is measured in units of the inverse lattice spacing.
The darkness of the shading is proportional to $S_{zz}(k,\omega)$.}
\label{fig3}
\end{figure}
 As we can see, a large part of the 
weight of the $j_E=0$ peak of the structure factor at $\pi$ shifts 
to $\pi\pm\delta k$ for $j_E\not=0$. Thus 
one can expect that even if $\delta k$ is small, the 
presence of a small energy current
will result in a {\em broadening} by $2\delta k$
of the inelastic peak centered at
wavevector $\pi$. It is this broadening that we propose as an
experimental signature for the current-carrying state.
The remaining question now is how to estimate $\delta k$.

As we can see from (\ref{kj}), an estimate of $\delta k$ 
requires the value of the energy current, $j_E$. 
Thus we should, in principle, calculate $j_E$ in a spin chain where 
the two ends are kept at different temperatures. We are unable 
to do this for any reasonable size system, and so we shall treat 
the energy flux as a parameter taken from experiments 
($j_E\equiv j_E^{exp}$). Then a thermodynamic measurement of
$j_E^{exp}$ can be used to estimate the value of $\delta k$ 
in an independent neutron-scattering experiment.
Below we shall show how to estimate $j_E^{exp}$
using parameters from a realistic experimental setup.

\begin{figure}[htb]
\vspace{3truecm}
\hspace{-1truecm}
\centerline{
        \epsfxsize=6cm
        \epsfbox{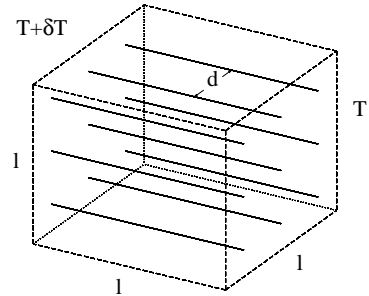}
           }
\vspace{-2.8truecm}
\caption{Experimental setup for measuring the effect of energy
flux. The solid lines represent the spin chains in a cubic
sample of volume $l^3$ with $d$ being the distance between the
chains. The energy flux is generated by keeping the two ends of 
the chains at temperatures $T$ and $T+\delta T$, respectively.}
\label{expsetup}
\end{figure}

Let the sample be a cube
of side $l=10^{-2}m$ and let the spin chains be 
along $x$ direction with the distance between 
the neighboring chains being $d = 10^{-9}m$.
Furthermore, let the sides of the cube perpendicular to the chains 
be at temperatures $T$ and $T+\delta T$ (see Fig.3). 
The temperature should be chosen to be low in order to minimize the phonon
contribution to $j_E^{exp}$. However, 
$T$ cannot be too small since then the small coupling between the chains 
makes the system three-dimensional.
The value of $\delta T$ should be, in principle, 
chosen large but the limitations 
of cryogenics of a realistic setup restrict the steady-state  
temperature-differences to $\delta T\le 0.1T$. From the above
considerations we arrive at the following ranges for the 
possible temperatures and temperature-differences 
 \begin{equation}
 T=(1-10)^oK \quad ; \quad \delta T =0.1T=(0.1-1)^oK \quad .
 \label{temps}
 \end{equation} 
The total current of heat across the sample can now be estimated as 
 \begin{eqnarray}
  j_E^{total}=\kappa l^2\frac{\delta T}{l}
 \label{Jheat}
 \end{eqnarray}
provided we know the heat conductivity, $\kappa$. We note here that 
the finiteness of the experimental $\kappa$ is not in contradiction 
with the singular nature ($\kappa^{int}=\infty$) of the heat conductivity 
of integrable spin chains. A macroscopic sample 
consist of spin chains of characteristic length $\ell \approx10^{-4}cm$
and the energy must also be transported between chains. This leads 
to the loss of ideal conductivity and results in a finite $\kappa$. 
Consequently, the estimate of 
energy flux using the experimental $\kappa$ does give an estimate of the 
energy flux through the chains provided the spin chains are 
the main channels of energy transport.

Unfortunately, $\kappa$, is not available for the materials we have in mind,
\cite{{kcuf3},{cs2cocl4},{cubenz},{cscucl}}, 
and another problematic issue is how much 
of the conductivity comes from the spin-chains.
Since measurements of the magnetothermal conductivity 
of magnetic materials 
\cite{{heatcond1},{heatcond2}} indicate that 
spin waves provide a significant fraction of the low-temperature 
thermal conductivity, we shall assume that an order of magnitude 
estimate of the energy current through the spin chains is
given by $j_E^{total}$. Furthermore, we shall assume that,
as a value of $\kappa$, we can take a characteristic value of 
this parameter in crystalline magnetic 
materials in the temperature range $T=(1-10)^oK$ \cite{heatcond.exp}:
 \begin{equation}
  \kappa \approx 
  (1-10)\cdot \frac{W}{m\cdot {}^oK}\quad .
 \label{kappa}
 \end{equation} 
We can then estimate $j_E^{total}\approx (10^{-3}-10^{-1})W$ and,
since the number of spin-chains in the sample is 
${\cal N}=l^2/d^2=(10^{-2}/10^{-9})^2=10^{14}$, we 
obtain the energy flux per chain, $j_E^{exp}$, as 
 \begin{equation}
  j_E^{exp}\approx\frac{{j_E^{total}}}{{\cal N}}=
  (10^{-17}-10^{-15}) W\quad .
 \label{jQ}
 \end{equation}
As we have seen (\ref{deltak}) 
the natural unit of energy current in a spin chain
is $j_E^{(1)}=J^2/h$. Using a characteristic value of $J=(1-10)^oK$ for the 
spin coupling we find $j_E^{(1)}\approx (10^{-12}-10^{-10})W$ and obtain the following 
estimate for the shift of the wavenumber
\begin{equation}
\delta k \sim \sqrt{\frac{j_E^{exp}}{j_E^{(1)}}} \sim 10^{-4}-10^{-2}\quad .
\label{deltakest}
\end{equation}
This is our central result. Since $\delta k \sim 10^{-2}$ is accessible 
in an inelastic neutron scattering experiment, the effect of shift in the
wavenumber should be observable. 

In summary, we have studied an integrable system which doesn't obey 
Fourier's law. We proposed that, under some simplifying assumptions, 
one can explore states of this system which carry current of energy
and, furthermore, one can derive 
theoretical results verifiable in experiments.


\section*{Acknowledgement}
I thank T. Antal, R. Cowley, F. Essler, A. R\'akos, L. Sasv\'ari, 
G. M. Sch\"utz, and A. Tsvelik for helpful discussions.
This work has been supported by the 
by the Hungarian Academy of Sciences (Grant No. OTKA T 029792).


\end{multicols}


\begin{references}
\bibitem{notehxx} The correlation function 
                  $\langle \sigma_n^z(t)\sigma_0^z(0)\rangle$ can be calculated 
                  equally easily for the case of time evolution by 
                  $\hat H_{\rm xx}-\lambda \hat j^E$. We believe, however,
                  that this is not the quantity that is measured 
                  in neutron scattering experiments.
\end{references}
\end{document}